\documentclass[aps,floats,twocolumn,superscriptaddress,nobalancelastpage,prb,showpacs]{revtex4}
\usepackage[final]{graphicx}
\DeclareGraphicsExtensions{.eps,.ps,.pdf}
\usepackage{amsmath,amssymb,bbold,bm}
\usepackage{multirow}

\begin{document}

%%%%%%%%%%%%%%%%%%%%%%%%%%% TITLE

\title{
NMR study of electronic correlations in Mn-doped Ba(Fe$_{1-x}$Co$_x$)$_2$As$_2$ and  BaFe$_{2}$(As$_{1-x}$P$_x$)$_2$
}

%%%%%%%%%%%%%%%%%%% AUTHORS

\author{D. LeBoeuf}
%\altaffiliation{david.le-boeuf@u-psud.fr}
\affiliation{Laboratoire de Physique des Solides, Univ. Paris-Sud, UMR8502, CNRS - F-91405 Orsay Cedex, France, EU}

\author{Y. Texier}
\affiliation{Laboratoire de Physique des Solides, Univ. Paris-Sud, UMR8502, CNRS - F-91405 Orsay Cedex, France, EU}

\author{M. Boselli}
\affiliation{Laboratoire de Physique des Solides, Univ. Paris-Sud, UMR8502, CNRS - F-91405 Orsay Cedex, France, EU}
\affiliation{Department of Physics, ``A. Volta'' University of Pavia-CNISM, I-27100 Pavia, Italy, EU}

\author{A. Forget}
\affiliation{Service de Physique de l'\'Etat Condens\'e, Orme des Merisiers, CEA Saclay (CNRS URA 2464), 91191 Gif sur Yvette Cedex, France, EU}

%\author{F. Rullier-Albenque}
%\affiliation{Service de Physique de l'\'Etat Condens\'e, Orme des Merisiers, CEA Saclay (CNRS URA 2464), 91191 Gif sur Yvette Cedex, France, EU}

\author{D. Colson}
\affiliation{Service de Physique de l'\'Etat Condens\'e, Orme des Merisiers, CEA Saclay (CNRS URA 2464), 91191 Gif sur Yvette Cedex, France, EU}

\author{J. Bobroff}
\affiliation{Laboratoire de Physique des Solides, Univ. Paris-Sud, UMR8502, CNRS - F-91405 Orsay Cedex, France, EU}

\date{\today}

%%%%%%%%%%%%%%%%%%%%%%%%%%%% ABSTRACT

\begin{abstract}

We probe the real space electronic response to a local magnetic impurity in isovalent and heterovalent doped BaFe$_2$As$_2$ (122) using Nuclear Magnetic Resonance (NMR). The  local moments carried by Mn impurities doped into Ba(Fe$_{1-x}$Co$_x$)$_2$As$_2$ (Co-122) and  BaFe$_{2}$(As$_{1-x}$P$_x$)$_2$ (P-122) at optimal doping induce a spin polarization in the vicinity of the impurity. The amplitude, shape and extension of this polarisation is given by the real part of the susceptibility $\chi'(r)$ of FeAs layers, and is consequently related to the nature and strength of the electronic correlations present in the system. We study this polarisation using $^{75}$As NMR in Co-122 and both  $^{75}$As and  $^{31}$P NMR in P-122. The NMR spectra of Mn-doped materials is made of two essential features. First is a satellite line associated with nuclei located as nearest neighbor of Mn impurities. The analysis of the temperature dependence of the shift of this satellite line shows that Mn local moments behave as isolated Curie moments. The second feature is a temperature dependent broadening of the central line. We show that the broadening of the central line follows the susceptibility of Mn local moments, as expected from typical RKKY-like interactions. This demonstrates that the susceptibility $\chi'(r)$ of FeAs layers does not make significant contribution to the temperature dependent broadening of the central line. $\chi'(r)$ is consequently only weakly temperature dependent in optimally doped Co-122 and P-122. This behaviour is in contrast with that of strongly correlated materials such as underdoped cuprate high-$T_{\rm c}$ superconductors where the central line broadens faster than the impurity susceptibility grows, because of the development of strong magnetic correlations when $T$ is lowered. Moreover, the FeAs layer susceptibility is found quantitatively similar in both heterovalent doped and isolvalent doped BaFe$_2$As$_2$.

%Ba(Fe$_{1-x-y}$Co$_x$Mn$_{y}$)$_2$As$_2$ and  Ba(Fe$_{2-y}$Mn$_y$)$_{2}$(As$_{1-x}$P$_x$)$_2$ around optimal doping.

\end{abstract}

\maketitle

%%%%%%%%%%%%%%%%%%%%%%%%%%%% I. INTRODUCTION

\section{Introduction}

Superconductivity in Fe-based materials can be induced in a remarkable amount of fashions \cite{paglione:natphy10}: carrier doping\cite{rotter08,sefat08} (heterovalent substitution), chemical pressure\cite{jiang09,sharma10} (isovalent substitution) and hydrostatic pressure\cite{alireza:jphycondmat09}. In each case the experimental phase diagrams are very similar: magnetism becomes unstable as a function of the tuning parameter and superconductivity arises around the magnetic instability. There is nonetheless a rich palette of variations from one material to the other, and physical properties such as superconducting gap structure, superconducting critical temperature $T_{\rm c}$, Fermi surface, magnetic ordered moments may vary significantly \cite{johnston:advphy10}. Such variations appear as well in the strength of electronic correlations present in those materials. Different experimental signatures of those electronic correlations have been reported, such as reduced kinetic energies \cite{qazilbash:natphy09}, enhanced masses \cite{walmsley:prl13}, %bandwidth narrowing \cite{borisenko:prl10},
 deviations from conventionnal $(T_1T)^{-1}$ Korringa law due to the presence of AF spin fluctutations\cite{nakai:prl10,ning:prl10}. The strength of electronic correlations assessed from those experimental findings vary significantly from one material to another\cite{yin:natmat11}, and whereas mass enhancement ($m^\star/m_{\rm band}$) can reach 24 in KFe$_2$As$_2$\cite{terashima:jpsj10}, it can be as low as 2 in LaFePO\cite{coldea:prl08}. %(Note that mass enhancement may as well vary significantly from one band to another within one material).

The experimental observation of those variations raises a central question in the field : can those systems fall under a universal description? The answer is not clear so far and basic issues about how strong are correlations within each family or iron-based superconductors%\cite{tesanovic:physics09}
, and how correlation strength may vary from one family to another %how different correlation strength can account for the observed variation in physical properties, or finally can different tuning mechanisms generate different correlation strength, 
are still not completly resolved. In order to determine the importance of electronic correlations in the description of those systems, it is necessary to perform experiments sensitive to correlation strength on different families of iron-based superconductors, and to compare those experiments with well established examples of materials where correlations are significant.

Here we use NMR to study the strength of magnetic correlations in two particular BaFe${_2}$As$_{2}$~(122) materials at optimal doping: heterovalent doped Ba(Fe$_{1-x}$Co$_x$)$_2$As$_2$ (Co-122) with $x_{\rm Co}\simeq6.5$\% and isovalent doped BaFe$_{2}$(As$_{1-x}$P$_x$)$_2$ (P-122) with $x_{\rm P}= 35$\%. To do this we introduce diluted Mn impurities that substitute to Fe. It has been shown that in Mn-doped BaFe${_2}$As$_{2}$, that the hole carried by Mn does not delocalise but rather acts as a local moment \cite{texier:epl12}. This impurity local moment induces a staggered spin polarization in its vicinity. This phenomenon has been studied extensively in metals (RKKY interaction) and correlated materials\cite{alloul:rmp09}. The impurity induced staggered polarization can be probed using the NMR of nuclei neighboring these local moments. The polarisation results in the appearance of satellites on the sides of the central line and in a temperature dependent broadening $\Delta \nu(T)$~of the NMR spectrum which in the diluted regime is:
\begin{equation}
\Delta\nu(T)\propto \langle S_{\rm imp}(T)\rangle \cdot  y_{\rm imp} \cdot f(\chi'_{\rm syst}(r,T)) \label{eq:main}
\end{equation} where $\langle S_{\rm imp}(T)\rangle$ is the thermodynamic average of the impurity spin, $y_{\rm imp}$ the impurity concentration and $ f(\chi'_{\rm syst}(r,T))$ a function of the real part of the real space susceptibility of the system, which determines the amplitude, the extension and the shape of the staggered spin polarization induced around the impurity. Knowing the temperature dependence of the susceptibility of the impurity local moments $\langle S_{\rm imp}(T)\rangle$, one can access the $T$-dependence of $\chi'_{\rm syst}(r)$. In a metal, it has been shown that RKKY interaction leads to a Lorentzian broadening $\Delta \nu(T)$ that follows the $T$-dependence of the local moment susceptibility, because $\chi'_{\rm syst}(r)$ is mainly a constant as a function of $T$ \cite{walstedt:prb74}. However in correlated systems, $\Delta \nu(T)$ can have a different $T$- dependence, due to the existence of magnetic correlations that imprint a $T$-dependence to $\chi'_{\rm syst}(r)$\cite{bobroff:prl97}. Assuming that the presence of the impurity does not locally affect the nature of the magnetic interactions, the NMR response in the vicinity of a local moment is consequently a good probe of magnetic correlations in a system.

Here we show that in the case of Mn-doped Co-122 and P-122, the broadening of the central line follows the susceptibility of Mn local moments. This demonstrates that the susceptibility of FeAs layers is only weakly temperature dependent in both materials. This behaviour is in contrast with that of the underdoped cuprate superconductor YBa$_2$Cu$_2$O$_y$, where the CuO$_2$ layer susceptibility has a strong $1/T$ dependence\cite{bobroff:prl97}.

The Article is organized as follows. 
In sec.~II, we describe the samples studied here and the measurement technique used.
In sec.~III, we present the evolution of the NMR spectra as a function of Mn doping in both Co-122 and P-122 materials.
In sec.~IV, we determine the susceptibility of Mn impurities local moments.
In sec.~V, we show that the broadening of the central line follows the susceptibility of Mn local moments, and from that we conclude that the susceptibiltiy of FeAs plane in Co-122 and P-122 is only weakly temperature dependent.
In sec.~VI, we finally discuss the implications of this result and compare it with other probes of correlation strength in BaFe${_2}$As$_{2}$.

%%%%%%%%%%%%%%%%%%%%%%%  II. METHODS

\section{Samples and Methods}

\begin{table}[h]
	\begin{center}
		\begin{tabular}{c|c|c|c}
\hline

Material & $x$ (\%) & $y_{\rm Mn}$ (\%)  & $T_{\rm c}$ (K)\\
\hline
\hline
\multirow{4}{*}{ Ba(Fe$_{1-x-y}$Co$_{x}$Mn$_{y}$)$_2$As$_2$} & 6.5 & 0.65 & 18 \\
 &7.5& 0.9 & 13.5 \\
 &6.0& 1 & 12.5 \\
 &6.5& 1.5 & 8 \\
\hline
\multirow{3}{*}{Ba(Fe$_{2-y}$Mn$_y$)$_{2}$(As$_{1-x}$P$_{x}$)$_2$} & 35& 0 & 30 \\ 
&35 & 1.5 & 16 \\
 &35 & 3.0  & $<2$\\
\hline
	\end{tabular}
	\end{center}
\caption{Characteristics of the samples : Mn contents and superconducting transition temperatures $T_{\rm c}$ for both Co-Mn-122 and P-Mn-122. In Co-Mn-122, the growth technique described in the text yields homogeneous Mn content amongst different samples from the same batch, except for $y_{\rm Mn}=0.65$ where a variation $y_{\rm Mn}=0.65\pm0.15$ was observed.
}
\label{tab:sample}
\end{table}

We studied 4 single crystals of Ba(Fe$_{1-x-y}$Co$_x$Mn$_{y}$)$_2$As$_2$ (Co-Mn-122) with Co content $x_{\rm Co}\sim 6.5$\% and Mn doping ranging from $y_{\rm Mn}=0.65$\% to $y_{\rm Mn}=1.5$\% (see table \ref{tab:sample}). The crystals were grown using a self-flux method from a mixture of Ba and FeAs, CoAs and MnAs powders in the molar ratio 1:4-x-y:x:y. The mixtures were heated in evacuated quartz tubes at $1180^\circ$C for 6 hours, cooled down to $1000^\circ$C at $6 ^\circ$C/hour, then down to room temperature. Clean crystals of typical dimensions $5\times5\times0.05 \rm~mm^3$ were mechanically extracted from the solid flux. Chemical analyses were performed with an electron probe (Camebax 50) on several crystals for each Co and Mn doping. SQUID measurements show that the superconducting transition temperature $T_{\rm c}$ decreases by $\sim-9$ K for every percent of Mn (see table \ref{tab:sample}), while staying well defined for all dopings. For the sake of clarity, most of the data presented here focus on samples with Mn doping  $y_{\rm Mn}=0.65$\% and  $y_{\rm Mn}=1.5$\%, though nearly identical results were obtained with the other samples. 

%%%%%%%%%%%%%%%%%%% FIGURE 2
%%%%figure/Marghe/SpettrivsT_0_1.5_3%_paper
%%%% inkscape E:\Users\leboeuf\work\papier\Figure\Data\Fig2_big
\begin{figure*}[t]
%\begin{figure}[t]
\center
	\includegraphics[width=18cm]{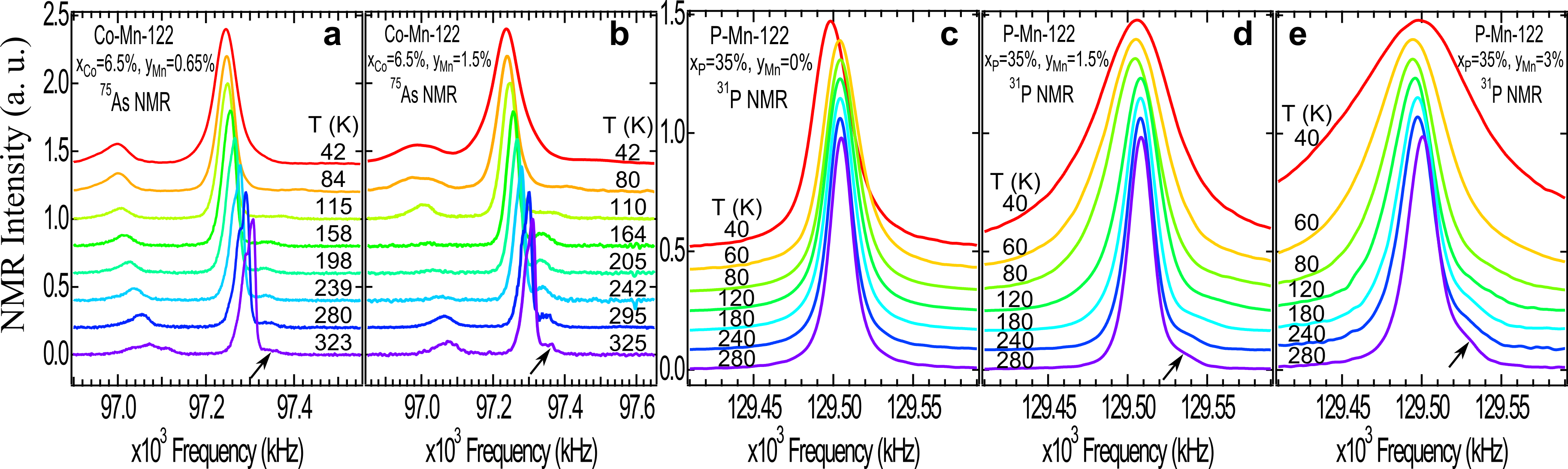}
	\caption{(Color online) Pannels a and b: $^{75}$As NMR measured with $H_0=13.3$ T, $H_0\parallel c$,  in Co-Mn-122 single crystals with $x_{\rm Co}=6.5$\% $y_{\rm Mn}=0.65$\% (pannel a) and $y_{\rm Mn}=1.5$\% (pannel b). The main line develops a structure at high temperature indicating that it is composed of several peaks with similar intensities. This structure might be a consequence of a modification of the hyperfine coupling or quadrupolar environment due to Co atoms. Pannel c, d and e: $^{31}$P NMR measured with $H_0=7.5$ T, $H_0\parallel c$, in P-Mn-122 powders with $x_{\rm P}=35$\%, $y_{\rm Mn}=0$ (pannel c), $y_{\rm Mn}=1.5$\% (pannel d), $y_{\rm Mn}=3.0$\% (pannel e). In pannels a,b,d and e, the arrow points to the satellite associated with As nuclei nearest neighbor to Mn atoms. In all pannels, curves are arbritarily shifted vertically for clarity.  
	}
	\label{fig:dataCo}
%\end{figure}
\end{figure*}

Ba(Fe$_{2-y}$Mn$_y$)$_{2}$(As$_{1-x}$P$_x$)$_2$ (P-Mn-122) powders with $x_{\rm P}=35$\%  and with Mn doping $y_{\rm Mn}=0$\%, $y_{\rm Mn}=1.5$\% and $y_{\rm Mn}=3.0$\% were synthesized using a solid state technique at $950^\circ$C during 36 hours from a stoichiometric mixture of Ba lumps, Fe, MnAs, P and As powders. The MnAs powder was prepared from a stoichiometric mixture of Mn and As powders heated in an evacuated quartz tube at $520^\circ$C for 10 hours and then $870^\circ$C for 10 hours. X-ray powder diffraction revealed a small amount of Fe$_2$P and a minority phase of an unidentified iron based compound. Thanks to the ability of NMR to adress and resolve nuclei originating from different phases, the presence of parasitic phases in the powders does not constitute an obstacle for the present study and can be ignored. The powders were then mixed with a standard two component epoxy. Using the anisotropy of the susceptibility of iron pnictides and a rotating sample holder placed in a static 7.5 T magnetic field applied perpendicularly to the rotating axis of the sample holder, the crystals constituting the powder can be aligned while the epoxy hardens. As a result, the components of the powder are oriented such that their c-axis is perpendicular to the rotation axis. Success of the orientation procedure has been checked by measuring the angle dependence of the $^{75}$As and $^{31}$P NMR frequency when rotating the field with respect to the c-axis. $T_{\rm c}$ in P-Mn-122 decreases at a rate of $\sim-10$ K per \% of Mn, such that sample with $y_{\rm Mn}=3.0$\% shows no sign of superconducting transition down to 2 K (see table  \ref{tab:sample}).

$^{75}$As NMR in Ba(Fe$_{1-x-y}$Co$_x$Mn$_{y}$)$_2$As$_2$ and Ba(Fe$_{2-y}$Mn$_y$)$_{2}$(As$_{1-x}$P$_x$)$_2$ was performed with an applied magnetic field of 13.3 T. $^{31}$P NMR was measured at 7.5 T. All the spectra were acquired for field applied along the c-axis and using a standard Fourier transform recombination after a $\pi/2-\tau-\pi$ pulse sequence.

\section{Description of the NMR spectrum in doped B$\textrm{a}$F$\textrm{e}$${_2}$A$\textrm{s}$$_{2}$}

%\begin{figure}[h]
%\center
%\includegraphics[width=8cm]{Figure/Data/Co6p5Mn1p5.jpg}
%\caption{
%	}
%	\label{fig:dataCo}
%\end{figure}
%%%%%%%%%%%%%%%%%%%%%%%%%%%%

%%%%%%%%%%%%%%%%%%% FIGURE 1
%%%%%%%%%%%%% papier/figure/Marghe/spettrivsT_3%
%%%%%%%%%%%%%%%ùinkscape E:\Users\leboeuf\work\papier\Figure\Data\Fig1_sketch
\begin{figure}[t]
\center
	\includegraphics[width=8cm]{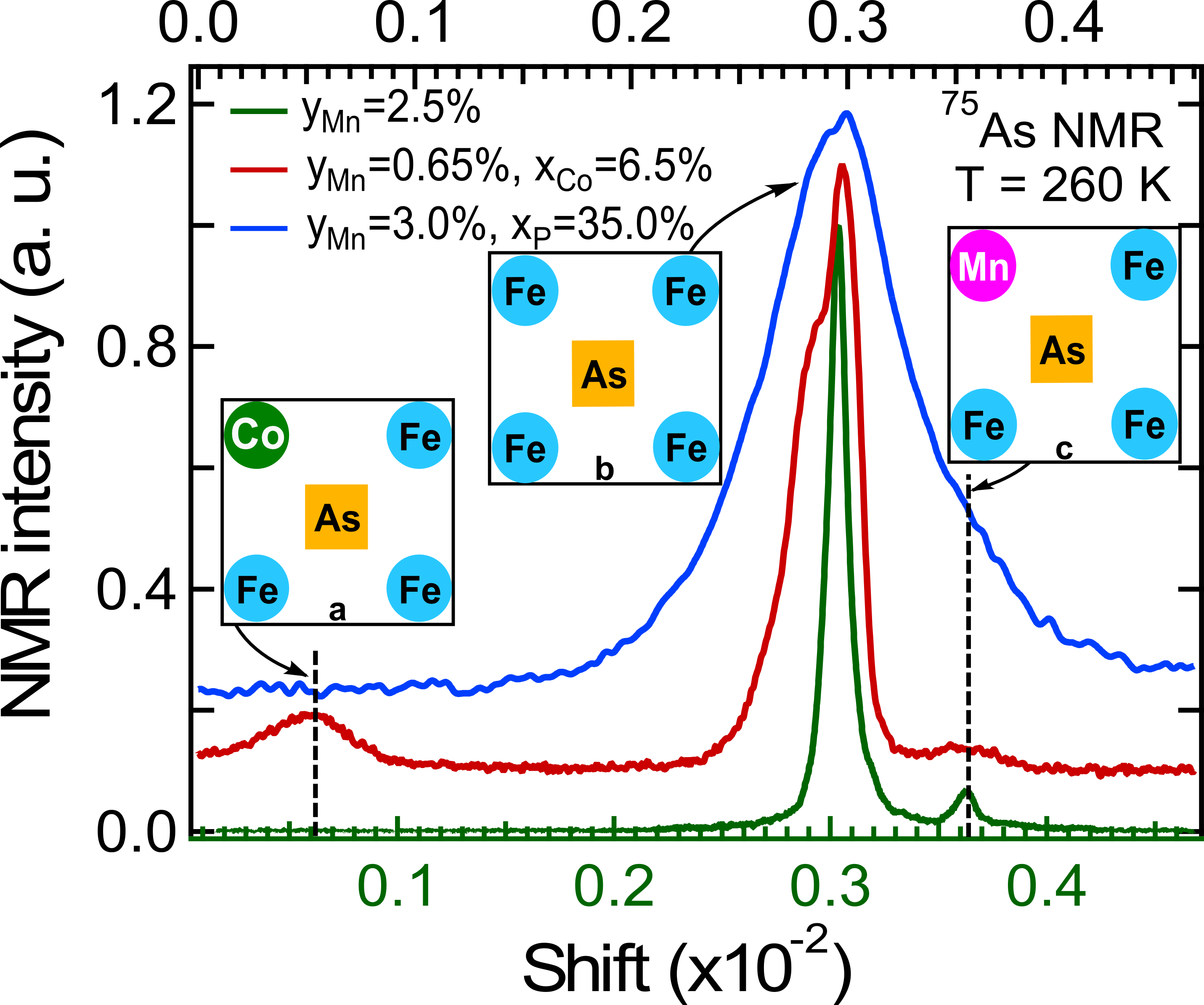}
	\caption{(Color online) Typical $^{75}$As NMR spectra at $T=260$ K in Mn-122 compounds, measured with $H_0=13.3$ T, $H_0\parallel c$. Pure Mn-122 ($y_{\rm Mn} = 2.5$\%) is shown in green (bottom axis) and features a satellite on the right wing of the central line associated with As nuclei nearest neighbor to Mn, as sketched in inset c (from ref.\cite{texier:epl12}). Red curve (top axis) shows Co-Mn-122 with $x_{\rm Co}=6.5$\%, $y_{\rm Mn}=0.65$\%. The satellite associated with Mn is still visible (right hand side of central line), and an additional satellite due to the modification of the second order quadrupole effect of As nuclei nearest neighbor to Co atoms (inset a) appears on the left wing of the central line. Blue curve (top axis) shows P-Mn-122 with $x_{\rm P}=35$\%, $y_{\rm Mn}=3.0$\%. The central line width has significantly increased in comparison to Co-Mn-122 because of higher disorder for $x_{\rm P}=35$\% than for $x_{\rm Co}=6.5$\%. The satellite associated with Mn is still here, but is merged with the central line. Curves are shifted vertically.
	}
	\label{fig:Asall}
\end{figure}

Figure~\ref{fig:dataCo} shows the temperature dependence of $^{75}$As NMR central line in Co-Mn-122 and $^{31}$P NMR central line in P-Mn-122, for different Mn concentrations. $^{75}$As NMR central line in Co-Mn-122 is splitted such that two additional peaks, associated with different local atomic environment, appear on each side of the main line (pannels a and b of figure~\ref{fig:dataCo}). The low frequency peak is due to a modification of the second order quadrupole effect in the vicinity of Co atoms and the high frequency peak (indicated by the arrows in fig.~\ref{fig:dataCo}a,b) comes from a modification of the shift in the vicinity of Mn impurities, as will be discussed below. $^{31}$P NMR central line in P-Mn-122 with $y_{\rm Mn}=0$ is not splitted (fig.~\ref{fig:dataCo}c).The introduction of Mn induces a splitting of the central line with a satellite appearing on the right hand side of the main line, as indicated by the arrows in fig.~\ref{fig:dataCo}d, e.

Let us first describe and explain in more detail the different features of the spectra shown in fig.~\ref{fig:dataCo}. We do so with the help of figure~\ref{fig:Asall} where we compare the $^{75}$As NMR central line in Mn doped materials: Mn-122 ( Ba(Fe$_{1-y}$Mn$_{y}$)$_2$As$_2$) (green curve), Co-Mn-122 (red curve) and P-Mn-122 (blue curve). Since $^{75}$As nuclear spin is $I=3/2$, its NMR frequency depends on both local magnetic field and  electric field gradients (EFG):
\begin{equation}
^{75}\nu=\frac{^{75}\gamma}{2\pi}H_0+\frac{^{75}\gamma}{2\pi}H_0 ~^{75}K+\alpha\frac{\nu_Q^2}{^{75}\gamma H_0}
\label{eq:nu75}
\end{equation} where $\nu_Q$ is proportionnal to the main axis EFG, $\alpha$ depends on the orientation of the crystal with respect to $H_0$ and on the asymmetry in the EFG and $^{75}\gamma/2\pi=7.2919$~MHz/T is the gyromagnetic factor of $^{75}$As nucleus. 

 $^{75}$K is the shift induced by the hyperfine interaction between the As nuclei and the Fe electrons orbitals and spin susceptibilities. The frequency of the main central line of the $^{75}$As NMR spectra originates from As nuclei located in regions far from any doping atom, a configuration sketched in fig.~\ref{fig:Asall}, inset b.  In 122 materials, where one As nucleus is surrounded by 4 Fe atoms, $^{75}$K is given by:
\begin{equation}
^{75}K_{\rm central}=4A_{\rm Fe}\cdot\chi'_{\rm Fe}+ K_{\rm orb}
\label{eq:k75}
\end{equation} with $\chi'_{\rm Fe}$ the Fe layer susceptibility probed far from any doping atom, $A_{\rm Fe}$ the hyperfine coupling between one Fe and the As nucleus, and $K_{\rm orb}$ a temperature independent shift due to orbital effects. 

In Mn-doped 122, the NMR frequency of As nuclei located as nearest neighbor of Mn atoms (a situation depicted in fig.~\ref{fig:Asall}, inset c) is modified. This results in the splitting of the NMR central line such that a satellite on the right-hand side of the main line appears (green curve in fig.~\ref{fig:Asall}). Eq.\eqref{eq:nu75} indicates that this satellite may arise from a modification of $^{75}$K or of the local EFG. A previous NMR study of Mn-122 has shown that the satellite shift is proportionnal to $H_0$ and that it is therefore due to a modification of the shift $^{75}$K (ref.\cite{texier:epl12}). 

By substituting Fe atoms with $x_{\rm Co}=6.5$\% Co atoms, we get Co-Mn-122 which typical $^{75}$As NMR spectrum is shown in red in figure \ref{fig:Asall}. Co substitution has three main consequences that we will discuss now. 
Eq.\eqref{eq:k75} indicates that $^{75}K_{\rm central}$ depends on carrier population in Fe orbitals. The first consequence of Co subsitution is thus a shift of the $^{75}$As NMR spectrum as observed in previous work\cite{ning:prl10}. The second consequence of Co substitution is the appearance of an additional satellite on the left-hand side of the central line, as shown in figures \ref{fig:dataCo}a, \ref{fig:dataCo}b and \ref{fig:Asall}. This satellite originates from the modification of the second order quadrupolar contribution to the NMR shift (the last term in eq. \eqref{eq:nu75}) in the vicinity of Co atoms, as shown earlier\cite{ning:prl10}. For symmetry reasons, $\alpha=0$ in the specific case of BaFe$_2$As$_2$, on the As atomic site, for $H_0\parallel c$, so that the position of the central line does not depend on the EFG. However Co substitution locally breaks this symmetry and shifts the frequency of As nuclei located as nearest neighbor of Co atoms (inset a in fig.~\ref{fig:Asall}), giving rise to this low frequency satellite (see figure \ref{fig:Asall}). This identification is confirmed by the fact that the $T$-dependence of the shift of this satellite follows the $T$-dependence of the sift of the central line (see fig.~\ref{fig:dataCo}a,b). The third consequence of Co doping, is a broadening of the NMR lines, due to increased structural and electronic disorder induced by Co-doping. This broadening is however small enough so that both the position of the satellite line and the full width at half maximum of the central line can be followed as a function of temperature (see fig.~\ref{fig:dataCo}a and b). Those quantities will be central for the discussion that follows in the next sections.

Finally, the blue curve in fig.~\ref{fig:Asall} is a typical  $^{75}$As NMR spectrum for P-Mn-122, with  $x_{\rm P}=35$\% and $y_{\rm Mn}=3$\%. Disorder is significantly increased in comparison to Co-Mn-122 with $x_{\rm Co}=6.5$\%. Now the high frequency Mn-related satellite is merged with the central line. Signature of this satellite is still slightly visible though in the asymmetry of the central line, with a higher weight on the right-hand side, where the satellite associated with As nuclei nearest neighbor to Mn atoms is expected. Note that there is no visible splitting due to second order quadrupolar effect. This can be a result of higher disorder than in Co-Mn-122, and/or of the fact that the As atomic sites are closer to Co atoms than to P atoms, meaning that this quadrupolar satellite could merge with the central line in the case of P-Mn-122.
The fact that the $^{75}$As NMR frequency is sensitive to EFG disorder results in a very broad line in the case P-Mn-122 sample, due to high P-doping level. This can be a major drawback in the study of the magnetic broadening of this line. To circumvent this problem we use $^{31}$P NMR in P-Mn-122. $^{31}$P has a nuclear spin $I=1/2$ and is hence insensitive to EFG :
\begin{equation}
^{31}\nu=\frac{^{31}\gamma}{2\pi}H_0+\frac{^{31}\gamma}{2\pi}H_0 ~^{31}K
\end{equation} with $^{31}\gamma/2\pi=17.2356$~MHz/T the gyromanetic factor of P nucleus.
Figures~\ref{fig:dataCo}c, \ref{fig:dataCo}d and \ref{fig:dataCo}e show the temperature dependence of $^{31}$P NMR in P-Mn-122 for three different Mn dopings. In the case of $y_{\rm Mn}=0$ and $x_{\rm P}=35.0$\%, the $^{31}$P NMR line width is $^{31}\Delta\nu\sim20$ kHz whereas the  $^{75}$As NMR line width in the same sample is $^{75}\Delta\nu\sim60$ kHz.  $^{31}$P NMR is thus a cleaner probe of magnetic braodening. The  $^{31}$P NMR line width observed here is a factor $\sim4.5$ smaller than that observed previously in mosaic of single crystals \cite{nakai:prl10}, demonstrating the quality of the powders and the efficiency of our orientation procedure.

In the following we will study the Mn-induced broadening of the $^{75}$As NMR central line in Co-Mn-122 and that of the $^{75}$As and $^{31}$P NMR central line in P-Mn-122. But in order to understand these broadenings we first need to determine the susceptibility of the local moments carried by Mn impurities.

%%%%%%%%%%%%%%%%%%% FIGURE 3

\begin{figure}[t]
\center
	\includegraphics[width=8cm]{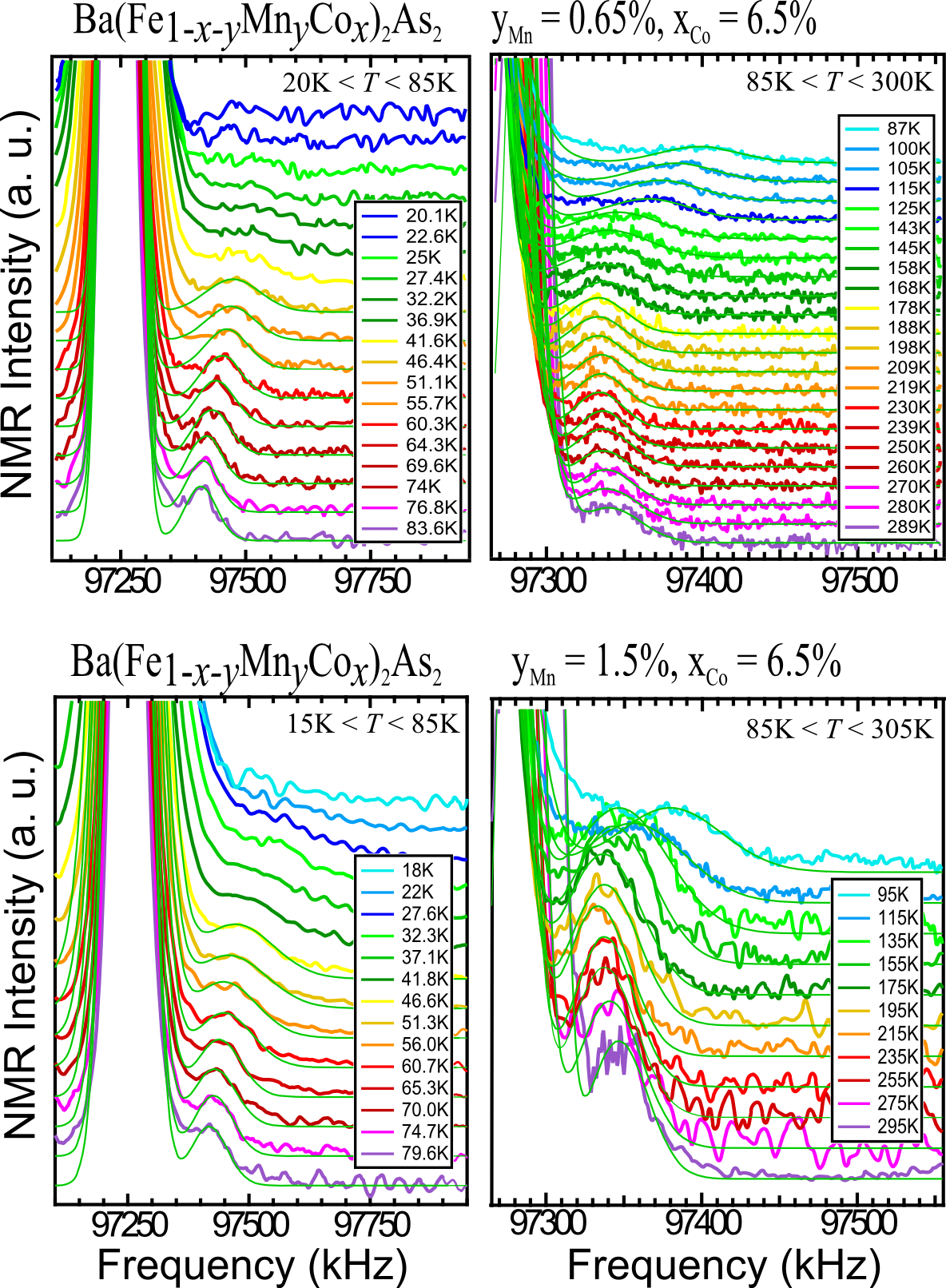}
	\caption{
	(Color online) Zoom on the $^{75}$As NMR satellite associated with As nearest neighbor to Mn atoms in Co-Mn-122 with $x_{\rm Co}=6.5$\%, for $15<T<300$ K, measured with $H_0=13.3$ T, $H_0\parallel c$. Green lines are fits to the experimental data used to determine the shift of satellite line $^{75}K_{\rm sat}$. Top pannels : $y_{\rm Mn}=0.65$\%. Bottom pannels : $y_{\rm Mn}=1.5$\%. In all cases, the satellite line merges with the main line for $T\lesssim50$~K, preventing us to confidently determine the shift of the satellite line in this temperature range.}
	\label{fig:Fitsat}
\end{figure}

%%%%%%%%%%%%% SECTION Chi Mn

\section{Determination of the M$\rm n$ local moment susceptibility}

%%%%%%%%%%%%%%%%%%% FIGURE 4

%%%%% Macro Chi Mn from E:\Users\leboeuf\work\yoan\yoan_CoMn\igor sauv\squid
%%%%% Chi Mn from NMR in igor file E:\Users\leboeuf\work\papier\Figure\FigV14_Yoth
\begin{figure}[t]
\center
	\includegraphics[width=8cm]{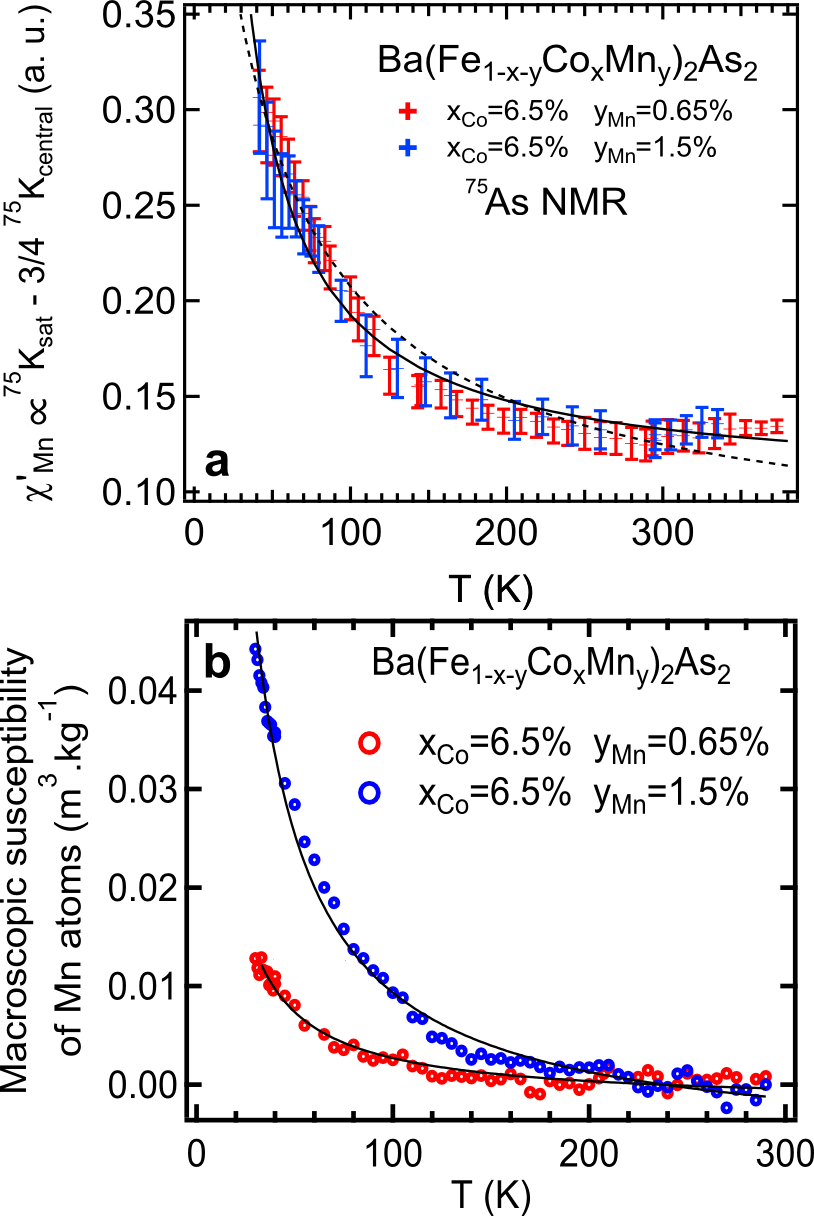}
	\caption{%fit $\chi_{\rm Mn}$ NMR : black line CW with $\theta=0~C=8.9515~y_0=0.103$, dashed line CW with $\theta=40~C=19.675~y_0=0.0669$
(Color online) Determination of Mn impurity local moments susceptibility $\chi'_{\rm Mn}$  in Co-Mn-122 ($x_{\rm Co}=6.5$\%) for $y_{\rm Mn}=0.65$\% (red) and  $y_{\rm Mn}=1.5$\% (blue). a) $\chi'_{\rm Mn}(T)$ determined with the $^{75}$As NMR shift. Continuous and dashed black lines are fits to the data with a Curie-Weiss model $y_0+C/(T+\theta)$ with $\theta$ respectively fixed to 0 and 40 K. 
b) $\chi'_{\rm Mn}(T)$ determined with macroscopic susceptibility measurements where the Fe layer contribution to the homogeneous susceptibility, namely $\chi'_{\rm Fe}(T) \propto ~^{75}K_{\rm central}(T)$ has been subtracted. Black lines are fits to the data with a Curie law $y_0+C/T$.
}
	\label{fig:chiMn}
\end{figure}

Texier \emph{et al.}\cite{texier:epl12} used the temperature dependence of the Mn satellite shift $^{75}K_{\rm sat}(T)$  in order to determine the susceptibility of the local moments carried by Mn impurties $\chi'_{\rm Mn}(T)$ in Mn-122. We apply the same method to our data on Co-Mn-122. In the diluted limit, each As site participating to the high-frequency satellite intensity is hyperfine coupled to 1 adjacent Mn atom and 3 Fe atoms (see inset c in fig.~\ref{fig:Asall}) such that eq.\eqref{eq:k75} is modified in the following way:
\begin{equation}
^{75}K_{\rm sat}=3A_{\rm Fe}\cdot(\chi'_{\rm Fe}+\Delta \chi')+A_{\rm Mn}\cdot\chi'_{\rm Mn} + K'_{\rm orb}
\end{equation} where $\chi'_{\rm Fe}$ is the Fe susceptibility far from Mn, and $\Delta \chi'$ accounts for possible modification of this susceptibility in the vicinity of Mn. Combining this with eq.\eqref{eq:k75}, assuming that $A_{\rm Fe}$ is unaffected and that $\Delta \chi' \propto \chi'_{\rm Mn}$ leads to: 
\begin{equation}
\chi'_{\rm Mn}(T)\propto~ ^{75}K_{\rm sat}(T)-\frac{3}{4}~^{75}K_{\rm central}(T) +K''_{\rm orb}
\end{equation} with $K''_{\rm orb}=3/4K_{\rm orb}-K'_{\rm orb}$, a $T$-independent orbital term. We first determine $^{75}K_{\rm sat}$ by fitting the spectra of fig. \ref{fig:dataCo}a,b. Fig.~\ref{fig:Fitsat} shows a zoom of the satellite and the fits to the spectra used to determine $^{75}K_{\rm sat}(T)$. $\chi'_{\rm Mn}(T)$ is then computed and plotted in fig.~\ref{fig:chiMn}a. The errorbars increase considerably when the temperature is lowered because both the satellite and the main line broaden faster than they shift apart. For $T\lesssim50$~K the satellite line merges with the main line such that the shift of the satellite line cannot be reliably determined anymore. The fact that curves for $y_{\rm Mn}=0.65$\% and $y_{\rm Mn}=1.5$\% overlap shows that the Mn induced polarisation does not depend on Mn concentration, and is evidence that As nuclei are coupled to at most 1 Mn atom. Additionnal confirmation that the Mn content is within the diluted regime comes from the Mn-doping dependence of the intensity ratio between the satellite and the central line, which follows the  probability law of finding As atoms with exactly one Mn nearest neighbor\cite{texier:epl12}. $\chi'_{\rm Mn}$ in Co-Mn-122 shows a similar temperature dependence to what was observed earlier in Mn-122\cite{texier:epl12}. A Curie-Weiss fit to $\chi'_{\rm Mn}(T)$ gives $\theta = 20 \pm 20$ K. Such a low value of $\theta$, and independent of Mn doping,  indicates that Mn-Mn interactions are weak.

This is further confirmed by an analysis of the macroscopic susceptibility of Co-Mn-122 samples measured by SQUID. By subtracting the Fe contribution to the macroscopic homogeneous susceptibility, namely $\chi'_{\rm Fe}(T) \propto ~^{75}K_{\rm central}(T)$, one can derive the Mn contribution to this macroscopic susceptibility (see Appendix A for details). The resulting macroscopic $\chi'_{\rm Mn}(T)$ is shown in fig.~\ref{fig:chiMn}b. The data fits nicely to a Curie model (black lines in fig.~\ref{fig:chiMn}b). A Curie-Weiss fit to those data would yield $\theta =0 \pm 10$ K. In a Curie model $\chi'(T)=y_0+C/T$, $C$ is proportionnal to $\mu_{\rm eff}^2$, with $\mu_{\rm eff}=g\sqrt{S(S+1)}$ the magnetic moment carried by the Mn impurity. Fits to the data of fig.~\ref{fig:chiMn}b give $\mu_{\rm eff}=1.63\pm0.3\mu_{\rm B}$ for all samples. This value is in agreement with $\mu_{\rm eff}=1.73$ calculated for a spin $S=1/2$ and $g=2$, as expected in a case where each Mn atom carries one localized hole.  

In summary, this study of the Mn local moment susceptibility is consistent with a scenario where each Mn atom carries a local $S=1/2$ moment, and where those local moments are only weakly coupled to each other, and behave as isolated Curie moments.

%%%%%%%%%%%%%%%%%%%%%%%%%%%% SECTION BROAD CENTRAL LINE

\section{Central line broadening}

%%%%%%%%%%%%%%%%%%% FIGURE 5
%%%%%%% Dnu for As in P-122 and Co-122 in igor exp.  E:\Users\leboeuf\work\papier\Figure\Fig_5.6.1_MB_DnuAs_CoMn_v2
%%%%%%% Dnu for P in igor exp :  figure/Marghe/Plots_FWHM_paper

\begin{figure}[h]
\center
	\includegraphics[width=8cm]{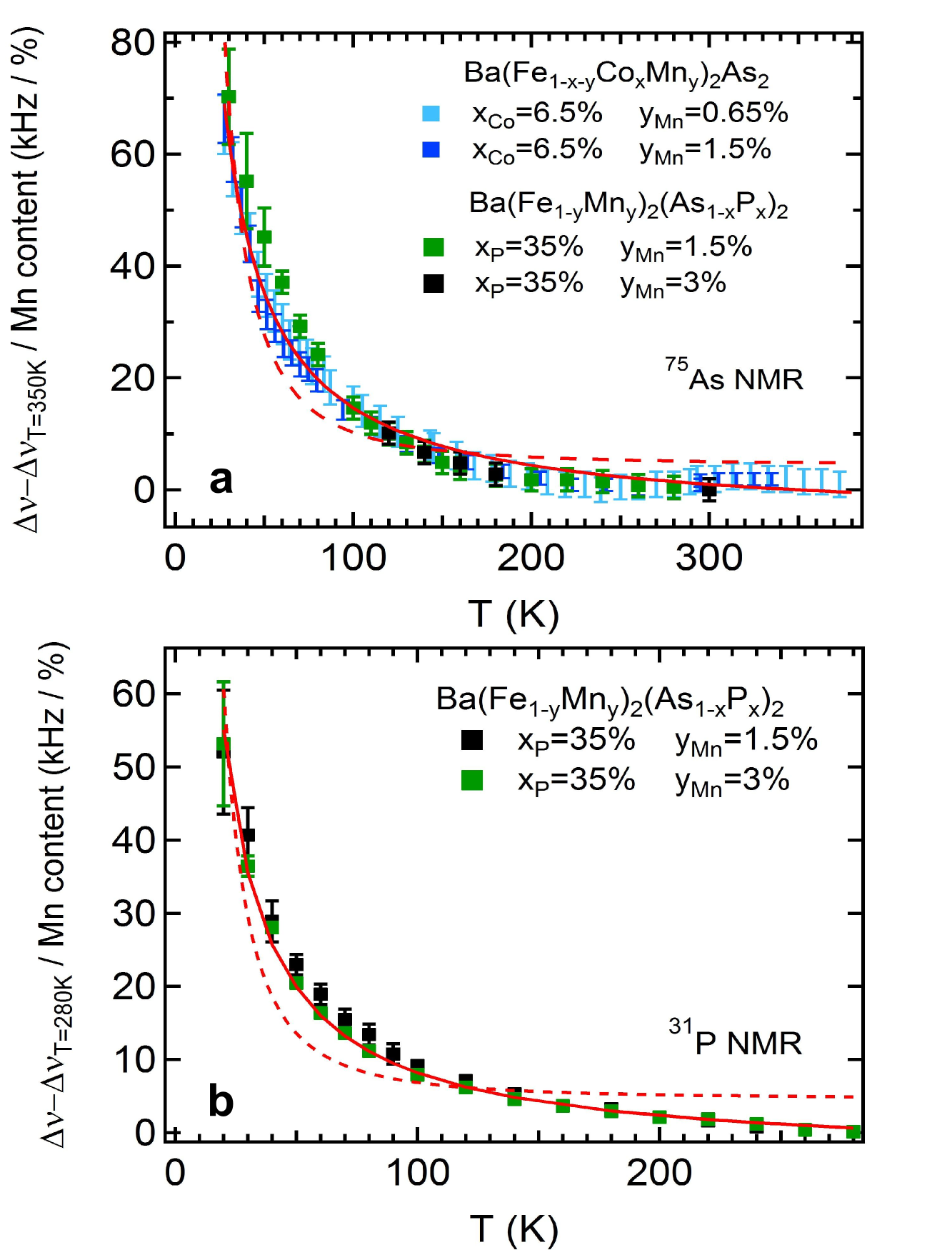}
	\caption{%$^{75}$As NMR : line CW fit with $\theta=0 (fixed),~y_0=-5.8878~C=2051.9$. Dash line $y_0+C/T^2$ fit with $y_0=4.3885,~C=58571$.
%$^{31}$P NMR : line CW fit with  $\theta=0 (fixed),~y_0=-3.539~C=1173$. Dash line $y_0+C/T^2$ fit with $y_0=4.6303,~C=22412$.
(Color online)  Central line broadening plotted as $(\Delta \nu-\Delta \nu_{\textrm{high}-T})/y_{\rm Mn}$. a) From $^{75}$As NMR spectra in both Co-Mn-122 (dark and light blue markers) and P-Mn-122 (black and green squares). b) From  $^{31}$P NMR spectra in P-Mn-122. In both pannels, red dashed line is a least square $y_0+C/T^2$ fit to the data and red continuous line is a least square Curie law $y_0+C/T$ fit to the data. While $^{75}$As NMR $\Delta\nu(T)$ is best reproduced by a Curie model in both cases. Incompatibility with the  $1/T^2$ fit is more obvious in the case of $\Delta\nu(T)$ deduced from $^{31}$P NMR in P-Mn-122. 
	}
	\label{fig:dnuCoP}
\end{figure}

%Shall we keep data above 300K? It fits better to the Curie law without. How is the shift for T$>$300K? It shifts so no problem with T. What about the shoulder structure on the left hand side of the central line?

The analysis of the shift of the NMR spectral weight originating from the population of $^{75}$As nuclei located as nearest neighbors to Mn impurities, namely the high frequency satellite in fig.~\ref{fig:dataCo}a,b and \ref{fig:Fitsat}, allowed to determine the $T$-dependence of the Mn local moments susceptibility. $^{75}$As nuclei located further away from Mn impurities do not contribute to the satellite amplitude, but rather to the central line broadening $\Delta \nu(T)$. In combination with $\chi'_{\rm Mn}(T)$, the determination of $\Delta \nu(T)$ allow to reveal the FeAs layer susceptibility $\chi'_{\rm Fe}(T,\mathbf{r})$  (see eq.\eqref{eq:main}), and the strength of magnetic correlations in Mn-doped Co-122 and P-122. 

We measure the full width at half maximum $\Delta \nu(T)$ of $^{75}$As NMR central line in Co-Mn-122 and of both $^{75}$As and $^{31}$P NMR central lines in P-Mn-122. In order to isolate the Mn-induced broadening from over sources of broadening, we subtract the line width measured at the highest $T$ (where the Mn impurity contribution to the line width becomes negligeable) to the measured $\Delta \nu(T)$. The resulting quantity, normalized by the Mn content $y_{\rm Mn}$, is plotted in figure~\ref{fig:dnuCoP}a in the case $^{75}$As NMR for Co-Mn-122 and P-Mn-122 and in figure~\ref{fig:dnuCoP}b in the case of $^{31}$P NMR for P-Mn-122. In each case, the normalized $\Delta \nu(T)$ obtained for different Mn contents overlap, confirming the linear dependence of $\Delta \nu(T)$ on $y_{\rm Mn}$ indicated in eq.\eqref{eq:main}. 

%%%%%%%%%%%%%%%%%%% FIGURE 6
%%%%Dnu vs Chi Mn from E:\Users\leboeuf\work\yoan\yoan_CoMn\igor sauv\CoMn_compil proc plotYo()
%%%% post production with inkscape figure/data/DnuvsChiMn_Co_AllMn.svg
\begin{figure}[t]
\center
	\includegraphics[width=8cm]{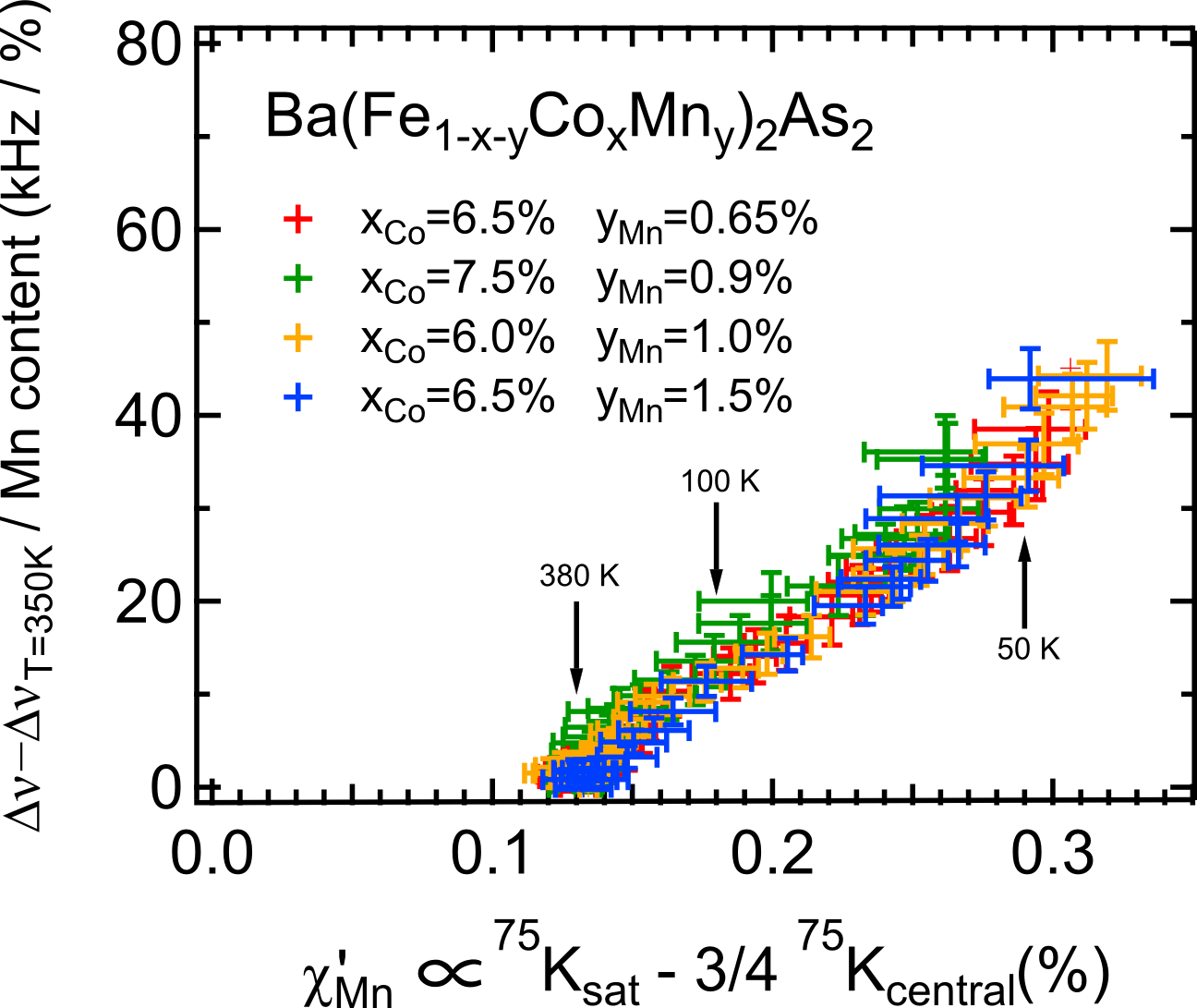}
	\caption{
(Color online) $^{75}$As NMRcentral line broadening plotted as a function of the Mn local moments susceptibility $\chi'_{\rm Mn}(T)$ derived from $^{75}$As NMR satellite shift, for all  Co-Mn-122 samples studied here. This plot shows that, within errorbars, $\Delta \nu(T) \propto \chi'_{\rm Mn}(T)$, such that the FeAs layer suscpetibility has a very weak $T$-dependence. There is remarkable agreement for all Co and Mn contents, indicating that Mn doping is within the diluted regime and that magnetic correlations are not dependent on carrier concentration within the errorbars.
	}
	\label{fig:dnuvschi}
\end{figure}

A Curie model nicely reproduces the temperature dependence of $\Delta \nu(T)$ in Co-Mn-122, as shown by the red line in fig.~\ref{fig:dnuCoP}a. It indicates that $\Delta \nu(T)$ follows the same $T$-dependence as $\chi'_{\rm Mn}(T)$. Figure~\ref{fig:dnuvschi} shows the central line broadening $\Delta \nu$ as a function of $\chi'_{\rm Mn}$ determined earlier in fig.~\ref{fig:chiMn}a. Within the errorbars, the two quantities track each other in a linear fashion. There is no strong deviation from proportionality between $\Delta \nu(T)$ and $\chi'_{\rm Mn}(T)$. This indicates that the magnetic susceptibility of FeAs layers $\chi'_{\rm Fe}(T)$ is only weakly temperature dependent in Co-Mn-122.

In P-Mn-122, we have not been able to determine  $\chi'_{\rm Mn}(T)$ via $K_{\rm sat}$ because the high frequency satellite is merged with the central line, even in $^{31}$P NMR spectra (see fig.~\ref{fig:dataCo}d,e). We assume that $\chi'_{\rm Mn}(T)$ in P-Mn-122 follows a Curie law $T$-dependence just as in Co-Mn-122. Two facts make this assumption reasonable. First, Mn acts as local moment in P-122, just as in Co-122 (no modification of the Knight shift as function of Mn doping, broadening of the central line, appearance of a satellite at the same position as in Co-Mn-122 in $^{75}$As NMR spectra, see fig.~\ref{fig:Asall}, as a consequence of the polarization around the Mn local moments). Second, Mn contents studied here in P-Mn-122 are within the same range of concentration studied in Co-Mn-122, such that the effects of Mn on the spectrum are also proportional to Mn content in the case of P-Mn-122 (ratio of the satellite intensity with respect to central line intensity, and central line broadening proportionnal to $y_{\rm Mn}$ as shown in fig.~\ref{fig:dnuCoP}a and b). 

Based on this assumption, the fact that $\Delta \nu(T)$ from $^{75}$As NMR in P-Mn-122 is quantitatively identitical to $\Delta \nu(T)$ from $^{75}$As NMR in Co-Mn-122, as plotted in fig.~\ref{fig:dnuCoP}a, shows that $\chi'_{\rm Fe}(T)$ is quantitatively similar in P-122 and Co-122. $\chi'_{\rm Fe}(T)$ is thus weakly $T$-dependent in both heterovalent and isovalent-doped 122. It also suggests that magnetic correlations have comparable strength in both materials such that $\chi'_{\rm Fe}(T)$ appears to be insensitive to the number of carriers in the system. $^{31}$P NMR can be used to confirm the temprature dependence of $\Delta \nu(T)$ in P-Mn-122. The absence of quadrupolar effect in $^{31}$P NMR makes it a cleaner probe of magnetism. As plotted in fig.~\ref{fig:dnuCoP}b, $^{31}$P NMR central line broadening follows a Curie model almost perfectly. $^{31}$P NMR shows more compelling evidence that  $\chi'_{\rm Fe}(T)$ is weakly $T$-dependent and strengthens the results derived from $^{75}$As NMR.

%%%%%%%%%%%%%%%%%%%%%%%%%%%%

\section{Discussion}

Studies of  magnetic correlations using the NMR of nuclei located in the vicinity of an impurity have been performed in high-$T_{\rm c}$ cuprate superconductors\cite{alloul:rmp09}. Those systems are doped antiferromagnetic (AF) charge-transfer insulators, and strong magnetic correlations extend over a wide range of doping. In the underdoped regime, the broadening of the NMR spectra induced by impurity local moments, is found to diverge as $1/T^2$. If $\chi_{\rm imp} \propto 1/T$, then according to eq.\eqref{eq:main}, it implies that the susceptibility of the CuO$_2$ plane, the basic common building block of cuprate superconductors, has a strong $1/T$ dependence. The staggered polarization induced by the impurity local moment grows faster than the impurity susceptibility itself. This behaviour is a consequence of the growth of significant AF correlations as a function of $T$, in underdoped cuprates. In the overdoped regime, the $1/T^2$ dependence of the broadening is lost and a more conventionnal $1/T$ law is recovered\cite{ouazi:prb04}.

In figure \ref{fig:dnuCoP} we compare $\Delta \nu(T)$ of both $^{75}$As NMR and $^{31}$P NMR with a fit of the form $y_0+C/T^2$ (dashed red line). Departure from this $1/T^2$ dependence is observed in both cases though more strikingly for $^{31}$P NMR. While a $1/T$ dependence nicely reproduces the data, a $1/T^2$ dependence is clearly incompatible with our data. We thus conclude that both isovalent and heterovalent doped 122 materials lie closer to a conventionnal metal or an overdoped cuprate than to an underdoped cuprate, in terms of correlation strength as inferred from this NMR technique.

Numerical simulations of the NMR spectrum of Co-Mn-122  were performed in order to determine the shape and the spatial extension of the staggered polarization in the vicinity of Mn impurities (see Appendix B). %A RKKY model, although strongly dependent on the input parameters, is not compatible with our observations since it systematically produces more than one satellite (see fig.~\ref{fig:simu}a). 
The best simulations are achieved when AF correlations are turned on, whether at $\mathbf{Q}_{\rm stripe}=(\frac{1}{2},\frac{1}{2})$ or $\mathbf{Q}_{\rm N\acute{e}el}=(1,0)$ (in r.l.u. with respect to the tetragonal structure). The staggered magnetization around the impurity rapidly decreases as a function of distance to the impurity, typically extending over $\lambda\sim$ 3-5 units of in-plane lattice parameter, or $\lambda=22\pm5$\AA. This characteristic length can be compared to some extent with the AF correlation length $\xi$ found in inelastic neutron scattering (INS). In Co-122, with $x_{\rm Co}=15$\%, the 9.5meV INS AF peak at $T=40$ K has a finite width that corresponds to $\xi=20\pm4$ \AA\cite{inosov:natphy10}, comparable to the spatial extension of the staggered polarization we measured here in real space by NMR.  Moreover, we found that for $y_{\rm Mn}=0.65$\% we can simulate the temperature dependence of the spectrum by simply considering the $1/T$ dependence of $\chi'_{\rm Mn}$ while keeping the spatial extension $\lambda$ constant and equal to 4 units of  in-plane lattice parameter (see fig.~\ref{fig:simu}b and \ref{fig:simu}c). This is consistent with our previous conclusion drawn from fig.~\ref{fig:dnuCoP}: magnetic correlations in the FeAs plane are only weakly $T$-dependent.

Short range, weakly temperature dependent AF correlations seem to be the origin of the $^{75}$As NMR central line broadening in Co-Mn-122. In fig.~\ref{fig:dnuCoP}a we see that $\Delta \nu(T)$ is almost identical in both Co-Mn-122 and P-Mn-122. This suggests that the very same short range, weakly $T$-dependent AF correlations exist in both heterovalent and isovalent doped BaFe$_2$As$_2$, around optimal doping. This similarity between the two systems adds up to an increasing list of others: similar maximum $T_{\rm c}$ (25 K for Co-122, 30 K for P-122), similar thermodynamic phase diagrams and similar coupling of magnetism and superconductivity to the lattice \cite{boehmer:prb12}. $(T_1T)^{-1}$ measured with $^{75}$As NMR in Co-122 is also found to be very similar to $(T_1T)^{-1}$ measured with $^{31}$P NMR in P-122 \cite{nakai:prl10,ning:prl10}. A remarkable resemblance between the two compounds and yet very different tuning parameters (chemical pressure vs carrier doping) and very different levels of disorder (as illustrated in fig.~\ref{fig:Asall} by the width of the $^{75}$As NMR central line). Authors of ref.\cite{boehmer:prb12} argue those similiraties might be related to the fact that Co and P doping act similarly on the Fe-As(P) bond length, which has been shown to be a central parameter to control $T_{\rm c}$ and to suppress magnetism (ref.\cite{boehmer:prb12,zinth:epl12} and ref. therein). Our results are consistent with such an explanation: they indeed suggest that AF correlations, which are believed to be the underlying architects of the phase diagrams of those systems, do not depend on charge carrier doping. Nonetheless there is at least one dissimilarity between the two systems: they have different superconducting gap structures. Indeed while Co-122 is fully gapped around optimal doping \cite{reid:prb10}, line nodes were found in the superconducting gap of optimally doped P-122 \cite{hashimoto:prb10}. Our results suggest that this difference in gap structure cannot be accounted for by a difference in magnetic correlation strength.
%, and might rather be related to differences in the Fermi surface of the two materials (cite systems where nodes disappear as a function of doping and where it is related to small change in FS). P-122 has to maintain compensation, Co-122 doesn't.

Finally we compare our results with other probes of correlation strength in 122 materials. Yang \emph{et al.}\cite{yang:prb09} performed x-ray absorption (XAS) and resonant inelastic x-ray scattering (RIXS) measurements in BaFe$_2$As$_2$ that showed remarkable similarity with Fe-based metals while showing only poor comparability with Fe-oxide insulators, highlitghing the itenerant character of iron-pnictides. They then extracted from their spectra a value of the Coulomb repulsion $U$, and Hund's coulping $J_{\rm Hund}$: $U\sim W >J_{\rm Hund}$, with $W$ the bandwidth, indicating the existence of only moderated electronic correlations. Qazilbash and coworkers\cite{qazilbash:natphy09} used the ratio of kinetic energy of the electron extracted from optical conductivity data $K_{\rm exp}$ to the kinetic energy computed from band theory $K_{\rm band}$. They found that the value of $K_{\rm exp}/K_{\rm band}$ in BaFe$_2$As$_2$ is similar to $K_{\rm exp}/K_{\rm band}$ in ovderdoped high-$T_{\rm c}$ cuprates. Our study of electronic correlation strength with NMR is in agreement whith those findings.

Probes of correlation strength in 122 materials indicate that they feature only moderated correlations, weaker than in underdoped cuprates superconductors. However, high value of Sommerfeld coefficient and mass enhancement of certain bands derived from quantum oscillations in KFe$_2$As$_2$ suggest electronic correlations may be stronger in the case of hole-doped 122 materials\cite{terashima:jpsj10,fukazawa:jpsj09,hardy:prl13}. Recent theoritical work proposed that hole-doped  KFe$_2$As$_2$ is more correlated than electron-doped Co-122 because it stands closer, in terms of band filling, to an orbital-selective Mott transition driven by Hund's coupling between the different orbitals at play\cite{deMedici12}. The significant variation of band renormalization from one band to another observed in  KFe$_2$As$_2$ is supporting evidence for proximity to this selective Mottness. This scenario calls for an experimental investigation of correlation strength in  KFe$_2$As$_2$, and comparisons with the work achieved here in Co-122 and P-122 using NMR might help quantify and understand the evolution of electronic correlations  in different iron-based superconductors.  

%(Due to its low scattering rate, P-122 appears to be a particularly well suited system for the study of quantum criticality. several evidences : 1/T1T, penetration depth, QO, heat capacity point toward the existence of a QCP inside the SC dome. How should that affect the magnetic correlations? Is our result consistent with a QCP?)

%%%%%%%%%%%%%%%%%% SUMMARY
\section{Summary}

We studied the strength of electronic correlations around optimal doping in both isovalent doped (P-122) and heterovalent doped (Co-122) BaFe$_2$As$_2$. To do so, we introduced local magnetic moments carried by diluted Mn impurities that polarise the FeAs plane in its vicinity. This polarization results in the appearence of a satellite and in a broadening of the NMR spectra that we measured with $^{75}$As NMR in Co-122 and with both $^{75}$As and $^{31}$P NMR in P-122. We used the satellite shift to demonstrate that the Mn impurity local moments behave as nearly isolated free moments. We then studied the $T$-dependence of the NMR line width and showed that it follows the $T$-dependence of the Mn-impurity susceptibility in both Co-122 and P-122. This behaviour indicates that magnetic correlations are only weakly $T$-dependent in both materials. Moreover, the quantitative agreement between the $T$-dependent NMR line broadening found in P-122 and Co-122 suggests that those magnetic correlations are independent of carrier doping. Numerical simulations of the  $^{75}$As NMR spectrum of Co-Mn-122 show that AF correlations, extending over 3-5 units of in-plane lattice parameters, reproduce satisfactorily the experimental data.

The $T$-dependence of the NMR line broadening observed here is in contrast with that observed in underdoped cuprate, where the CuO$_2$ layer susceptibility measured with NMR in the vicinity of impurities shows a significant $T$-dependence, due the development of strong AF correlations as a function of temperature. We conclude that both isovalent and heterovalent doped 122 materials lie closer to a conventionnal metal or an overdoped cuprate than to an underdoped cuprate, in terms of correlation strength as inferred from this NMR technique.

%%%%%%%%%%%%%%%%%%%%%% APPENDIX

\appendix
\section{Determination of the macroscopic susceptibility of  M$\rm n$ local moments}
%%%%%%%%%%%%%%%%%% FIG APPENDIX 

%%%%%%%%%%%%% inkscape Data/ChiMn_Appendixl.svg
%%%%%%%%%%%% from  E:\Users\leboeuf\work\yoan\yoan_CoMn\igor sauv\squid
\begin{figure}[h]
\center
	\includegraphics[width=8cm]{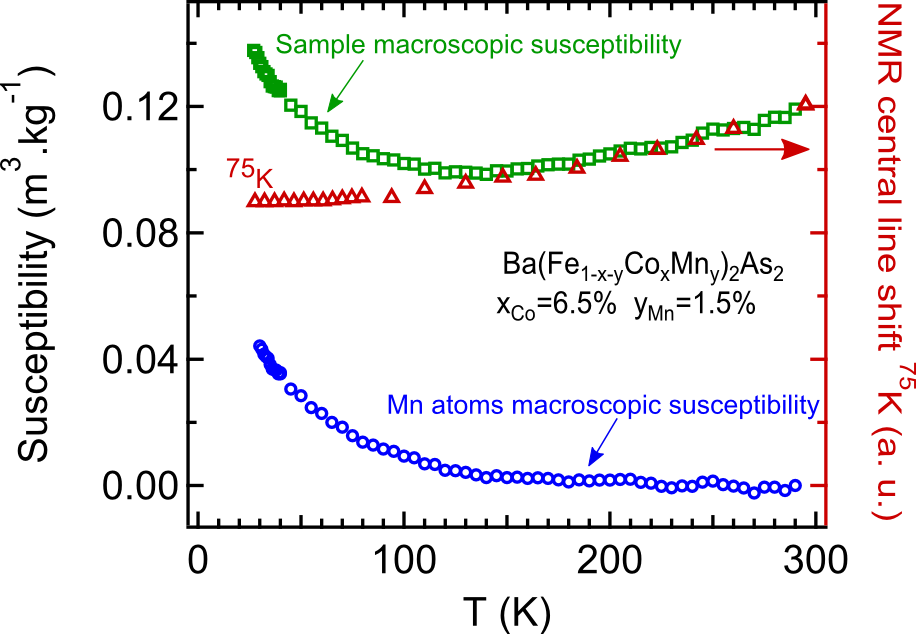}
	\caption{(Color online) Illustration of the method for the extraction of the macroscopic susceptibility of Mn local moments, using SQUID and NMR measurements. Green squares (left axis): sample (raw) macroscopic susceptibility obtained using the slope of $M(H)$ isotherms measured with SQUID. Blue circles (left axis): Mn local moments susceptibility contribution to the sample macroscopic susceptibility. It is obtained by subtracting the FeAs layer susceptibility, measured here with the shift $^{75}K$ of the central NMR line in the same sample of Co-Mn-122 (red triangles, right axis), to the sample macroscopic susceptibility (green squares).}
	\label{fig:chimnapp}
\end{figure}

Here we explain in more details the method we use to extract the Mn local moments contribution to the macroscopic susceptibility of Co-Mn-122. Magnetization measurements were performed up to 7 T using a Magnetic Properties Measurement System (MPMS) produced by Quantum Design. Isothermal $M(H)$ curves were measured and showed linear behaviour in the temperature range we focus on. Fits to $M(H)$ allow to extract the macroscopic susceptibility of the sample as a function of $T$, that is plotted in fig.~\ref{fig:chimnapp} (green squares) for Co-Mn-122 with  $x_{\rm Co}=6.5$\%, $y_{\rm Mn}=1.5$\%. The sample susceptibility is mainly the sum of two contributions: the FeAs layer susceptibility and the Mn local moments susceptibility. The FeAs layer susceptibility is measured in the same sample using NMR: it is given by $^{75}K(T)$, the shift of the central line of fig.~\ref{fig:dataCo}a,b, shown with red triangles in fig.~\ref{fig:chimnapp} for $x_{\rm Co}=6.5$\%, $y_{\rm Mn}=1.5$\%. By subtracting $^{75}K(T)$ (suitably multiplied by a factor such that $^{75}K(T)$ and the sample macroscopic susceptibility overlap for $T>200$~K) to the sample macroscopic susceptibility (green squares in fig.~\ref{fig:chimnapp}), we obtain the Mn local moments contribution, shown with blue circles in fig.~\ref{fig:chimnapp}.

\section{Simulations of the NMR spectrum}
%%%%%%%%%%%%%%%%%% FIG APPENDIX 

%%%%%%%%%%%%% inkscape Simu_2pannel.svg
%%%%%%%%%%%% from CoMn_simu_sat.svg and CoMn_simu_sp.svg in E:\Users\leboeuf\work\yoan\yoan_CoMn\figures_svg
\begin{figure}[h]
\center
	\includegraphics[width=8cm]{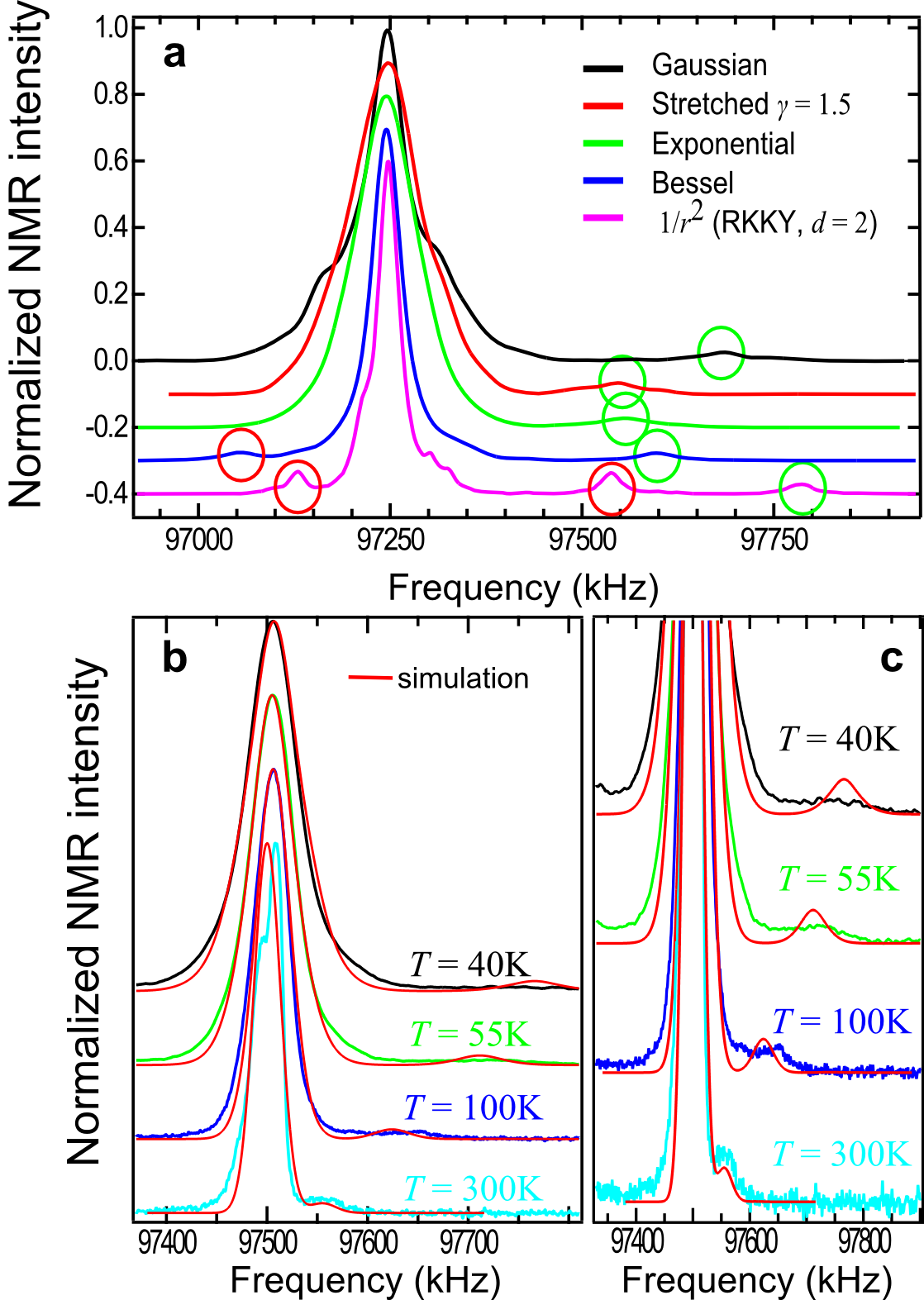}
	\caption{Simulations of  $^{75}$As NMR in Co-Mn-122 with $y_{\rm Mn}=0.65$\% and $x_{\rm Co}=6.5$\%. a) Simulated spectra obtained with different forms of polarisation (see text). Green circles highlight the satellite line which would be compatible with the experiment. Models giving rise to more than one satellite line (red circles) are not compatible with the experimental data. b) Experimental spectra ($H_0=13.3$ T, $H_0\parallel c$) at different temperatures compared with simulations obtained with $f_{\rm Stretched}$, $\gamma=1.5$, $Q=Q_{\rm stripe}$ and $\lambda=4$ (red lines). A zoom on the satellite line is shown in pannel c.
	}
	\label{fig:simu}
\end{figure}

We can elaborate to some extent about the shape and the spatial extension of the staggered magnetization induced in the vicinity of the Mn local moment by simulating the NMR spectrum. We do so in the case of $^{75}$As NMR in Co-Mn-122, following some of the methodology developed earlier in the case high-$T_{\rm c}$ cuprate YBCO\cite{ouazi:prb04}. The idea is to compare the simulated spectra given by a set of models and to find which model best reproduce the important features of the experimental data.

We simulate the NMR spectra under the assumptions that Mn concentration is within the diluted regime such that one As nucleus is nearest neighbor to at most one Mn atom, and such that the effect of several Mn atoms on the Fe electronic susceptibility is additive. Such assumption has been checked experimentally (see figures~\ref{fig:chiMn}, \ref{fig:dnuCoP} and  \ref{fig:dnuvschi}). In these simulations, we ignore the effect of Co substitution, since they mainly affect the local EFG environment of As nuclei. We assume that the hyperfine coupling between As nuclei and Fe atoms is isotrope. We compute the staggered magnetization in the FeAs layer and the resulting simulated spectra are convoluted with a Gaussian function to account for other sources of broadening that are already present in Co-122 without Mn doping.

We used two classes of model of the staggered magnetisation close to the impurity local moment, that we will discuss now.

\textbf{RKKY model}

The first class of model relies on a RKKY interaction. For a two-dimensionnal interaction, the staggered magnetization in the vicinity of the magnetic impurity takes the general form\cite{bealmonod:prb87}:
\begin{equation}
m_{\rm Fe}(k_{\rm F},r)\propto\frac{\textrm{cos}(2k_{\rm F}(r-1)+\phi)}{r^2} \label{eq:rkky}
\end{equation} with $k_{\rm F}$ the Fermi wavevector, $r$ the distance to the impurity and $\phi$ a phase we choose to be zero. Eq.\eqref{eq:rkky} is however not appropriate in the case of a multi-band system. The simulated spectra strongly depends on the value of $k_{\rm F}$ but systematically feature more than one satellite line, in disagreement with the experiments.

\textbf{Phenomenological models with AF correlations}

In this second class of model, we assume that the natural tendency of iron-pnictides towards the formation of AF order should appear in the shape of the staggered magnetisation such that:
\begin{equation}
m_{\rm Fe}(\mathbf{Q},\mathbf{r},\lambda)\propto\textrm{cos}(2\mathbf{Q_{\rm AF}}\cdot \mathbf{r})\cdot f(r,\lambda)
\end{equation} with $\mathbf{Q_{\rm AF}}$ the wavevector at which AF correlations are maximum, $\mathbf{r}$ the vector coordinate of each atomic site in a coordinate system where the Mn impurity seats at the origin, $\lambda$ a characteristic length controling the spatial extension of the staggered magnetization, and $f(r,\lambda)$ a function for the spatial damping of the staggered magnetisation.
We simulated NMR spectra with several $f(r,\lambda)$:
\begin{equation}
f_{\rm Bessel}(r,\lambda)=\frac{Y_k(0,r/\lambda)}{Y_k(0,1/\lambda)}
\end{equation} with $Y_k$ a Bessel function of the second kind.
\begin{equation}
f_{\rm Gaussian}(r,\lambda)=\textrm{exp}\Bigg({-\frac{r-1}{2\lambda / \sqrt{2\textrm{ln}2}}}\Bigg)^2
\end{equation} \begin{equation}
f_{\rm Stretched}(r,\lambda)=\textrm{exp}\Bigg({-\frac{r-1}{2\lambda / (2\textrm{ln}2)^{1/\gamma}}}\Bigg)^\gamma
\label{eq:stretch}
\end{equation} \begin{equation}
f_{\rm Exponential}(r,\lambda)=\textrm{exp}\Bigg({-\frac{r-1}{2\lambda / 2\textrm{ln}2}}\Bigg)
\end{equation}

In all the models mentionned above, we allow for a difference between the magnitude of the moment carried by Mn impurity and the moment carried by nearest neighbor Fe atoms.  

\textbf{Comparison to the data}

We now compare the different simulations we get from the different models of the staggered magnetisation we considered with the experimental spectra.
There are two main constraints that allow to eliminate models which are not compatible with the data: first is the number of satellite lines, second is the shape and width of the central line.
Fig.~\ref{fig:simu}a shows the simulated spectra for the different models exposed above. Amongst those models, three can already be discarded because they show more than one satellite line. Moreover, a simulation based on $f_{\rm Gaussian}$ do not yield a satisfactory shape for the central line. In the end, the model that best reproduces the experimental spectra is based on AF correlations at $\mathbf{Q}_{\rm stripe}$, with a damping function given by eq.\eqref{eq:stretch} with $\gamma=1.5$ and $\lambda\sim3-5$ units of in-plane lattice parameter. 

Using this particular model ($f_{\rm Stretched}$, $\gamma=1.5$, $\mathbf{Q}=\mathbf{Q}_{\rm stripe}$ and $\lambda=4$), and taking into account the $1/T$ dependence of $\chi'_{\rm Mn}(T)$, we were able to simulate the temperature dependence of the $^{75}$As NMR spectrum in Co-Mn-122 with $y_{\rm Mn}=0.65$\% and $x_{\rm Co}=6.5$\%, as shown in fig.~\ref{fig:simu}b,c. Both the width of the central line and the satellite shift of the experimental spectra are well reproduced by this specific model. 

These simulations do not however constitute a demonstration that the model we favour here ($f_{\rm Stretched}$, $\gamma=1.5$, $\mathbf{Q}=\mathbf{Q}_{\rm stripe}$ and $\lambda=4$) is the only model that could describe the data. Other models we did not consider here may work as well. Nonetheless, this approach allows to theorize the basic electronic correlations required to describe the NMR spectra, and since all models compatible with our data give a consistent value for $\lambda$, it also allows to constrain the spatial range of those correlations.

%%%%%%%%%%%%%%%%%%%%%%   REMERCIEMENTS

\begin{center}
%\centering
{\bf Acknowledgements}
\end{center}
The authors would like to thank H. Alloul, F. Bert, V. Brouet, P. Mendels, F. Rullier-Albenque for constructive discussions, S. Poissonnet (SRMP/CEA) for chemical analysis on the crystals 
 and A. Larcelet for help with the experimental setup. We acknowledge support from the French ANR PNICTIDES.
%

%%%%%%%%%%%%%%%%%%%%%%   REFERENCES


\begin{thebibliography}{99}

\bibitem{paglione:natphy10}
J. Paglione and R. L. Greene,  Nat. Phys. \textbf{6} 645 (2010)


\bibitem{rotter08}
M. Rotter, M. Tegel and D. Johrendt, Phys. Rev. Lett. \textbf{101} 107006 (2008)

\bibitem{sefat08}
A. S. Sefat, R. Jin, M. A. McGuire, B. C. Sales, D. J. Singh and D. Mandrus Phys. Rev. Lett. \textbf{101} 117004 (2008)

\bibitem{jiang09}
S. Jiang, H. Xing, G. Xuan, C. Wang, Z. Ren, C. Feng, J. Dai, Z. Xu and G. Cao,  J. Phys.: Condens. Matter. \textbf{21} 382203 (2009)

\bibitem{sharma10}
S Sharma, A. Bharathi, S. Chandra, V. R. Reddy, S. Paulraj, A. T. Saya, V. S. Sastry, A. Gupta and C. S. Sundar, Phys. Rev. B \textbf{81} 174512 (2010)

\bibitem{alireza:jphycondmat09}
P. L. Alizera, Y. T. Chris Ko, J. Gillett, C. M. Petrone, J. M. Cole, G. G. Lonzarich and S. E. Sebastian, J. Phys.: Condens. Matter \textbf{21} 012208 (2009)

\bibitem{johnston:advphy10}
D C. Johnston, Adv. Phys. \textbf{59} 803 (2010)

\bibitem{qazilbash:natphy09}
M. M. Qazilbash, J. J. Hamlin, R. E. Baumbach, Lijun Zhang, D. J. Singh, M. B. Maple and D. N. Basov, Nat. Phys. \textbf{5} 647 (2009)

\bibitem{walmsley:prl13}
P. Walmsley, C. Putzke, L. Malone, I. Guillamon, D. Vignolles, C. Proust, S. Badoux, A. I. Coldea, M. D. Watson, S. Kasahara, Y. Mizukami, T. Shibauchi, Y. Matsuda and A. Carrington, Phys. Rev. Lett. \textbf{110} 257002 (2013)

%\bibitem{borisenko:prl10}
%S. V. Borisenko, V. B. Zabolotnyy, D. V. Evtushinsky, T. K. Kim, I. V. Morozov, A. N. Yaresko, A. A. Kordyuk, G. Behr, A. Vasiliev, R. Follath abd B. B\"uchner, Phy. Rev. Lett. \textbf{105} 067002 (2010)

\bibitem{nakai:prl10}
Y. Nakai, T. Iye, S. Kitagawa, K. Ishida, H. Ikeda, S. Kasahara, H. Shishido, T. Shibauchi, Y. Matsuda and T. Terashima, Phys. Rev. Lett. \textbf{105} 107003 (2010)

\bibitem{ning:prl10}
F. L. Ning, K. Ahilan, T. Imai, A. S. Sefat, M. A. McGuire, B. C. Sales, D. Mandrus, P. Cheng, B. Shen and H.-H. Wen,  Phys. Rev. Lett. \textbf{104} 037001 (2010)

\bibitem{yin:natmat11}
Z. P. Yin, K. Haule and G. Kotliar, Nat. Mat. \textbf{10} 647 (2011)

\bibitem{terashima:jpsj10}
T. Terashima, M. Kimata, N. Kurita, H. Satsukawa, A. Harada, K. Hazama, M. Imai, A. Sato, K. Kihou, C.-H. Lee, H. Kito, H. Eisaki, A. Iyo, T. Saito, H. Fukazawa, Y. Kohori, H. Harima and S. Uji, J. Phys. Soc. Jpn. \textbf{79} 053702 (2010)

\bibitem{coldea:prl08}
A.I. Coldea, J.D. Fletcher, A. Carrington, J.G. Analytis, A.F. Bangura, J.-H. Chu, A.S. Erickson, I.R. Fisher, N.E. Hussey, and R.D. McDonald, Phys. Rev. Lett. \textbf{101} 216402 (2008) 

%\bibitem{tesanovic:physics09}
%Z. Tesanovic, Physics \textbf{2} 60 (2009)

\bibitem{texier:epl12}
 Y. Texier,  Y. Laplace, P. Mendels, J. T. Park,l G. Frieme, D. L.  Sun, D. S. Inosov, C. T. Lin and J. Bobroff, Europhys. Lett. \textbf{99} 17002 (2012)

\bibitem{alloul:rmp09}
H. Alloul, J.  Bobroff,  M. Gabay and P. J. Hirschfeld, Rev. Mod. Phys. \textbf{81} 45–108 (2009)
 
 \bibitem{walstedt:prb74}
R. E. Walstedt and L. R. Walker, Phys. Rev. B \textbf{9} 4857 (1974)

\bibitem{bobroff:prl97}
J. Bobroff, H. Alloul , Y. Yoshinari, A. Keren,  P. Mendels, N. Blanchard, G. Collin and J.-F. Marucco, Phys. Rev. Lett. \textbf{79} 2117 (1997)

%\bibitem{dorothee:prl09}
%F. Rullier-Albenque, D. Colson, A. Forget and H. Alloul, Phys. Rev. Lett \textbf{103} 057001 (2009)

\bibitem{ouazi:prb04}
S. Ouazi, J. Bobroff, H. Alloul and W. A. MacFarlane, Phys. Rev. B. \textbf{70} 104515 (2004)

\bibitem{inosov:natphy10}
D. S. Inosov, J. T. Park, P. Bourges, D. L. Sun, Y. Sidis, A. Schneidewind, K. Hradi, D. Haug, C. T. Lin, B. Keimer and V. Hinkov, Nat. Phys. textbf{6} 178 (2010)

\bibitem{boehmer:prb12}
A E. B\"ohmer, P. Burger, F. Hardy, T. Wolf, P. Schweiss, R. Fromknecht, H. v. L\"ohneysen, C. Meingast, H. K. Mak, R. Lortz, S. Kasahara, T. Terashima, T. Shibauchi and Y. Matsuda, Phys. Rev. B \textbf{86} 094521 (2012)

\bibitem{zinth:epl12}
V. Zinth and D. Johrendt, Europhys. Lett. \textbf{98} 57010 (2012)

\bibitem{reid:prb10}
J-P. Reid, M. A. Tanatar, X. G. Luo, H. Shakeripour, N. Doiron-Leyraud, N. Ni, S. L. Bud’ko , P. C. Canfield, R. Prozorov, and L. Taillefer, Phys. Rev. B \textbf{82} 064501 (2010)

\bibitem{hashimoto:prb10}
K. Hashimoto, M. Yamashita, S. Kasahara, Y. Senshu, N. Nakata, S. Tonegawa, K. Ikada, A. Serafin, A. Carrington, T. Terashima; H. Ikeda, T. Shibauchia and Y. Matsuda, Phys. Rev. B \textbf{81} 220501(R) (2010)

\bibitem{yang:prb09}
W L. Yang, A. P. Soni, C.-C. Chen, B. Moritz, W.-S. Lee, F. Vernay, P. Olalde-Velasco, J. D. Denlinger, B. Delley, J.-H. Chu, J. G. Analytis, I. R. Fisher, Z. A. Ren, J. Yang, W. Lu, Z. X. Zhao, J. van den Brink, Z. Hussain, Z.-X. Shen and T. P. Devereaux, Phys. Rev. B \textbf{80} 014508 (2009)

\bibitem{fukazawa:jpsj09}
H. Fukazawa, Y. Yamada, K. Kondo, T. Saito, Y. Kohori, K. Kuga, Y. Matsumoto, S. Nakatsuji, H. Kito, P. M. Shirage, K. Kihou, N. Takeshita, C. H. Lee, A. Iyo, and H. Eisaki, J. Phys. Soc. Jpn. \textbf{78} 083712 (2009)

\bibitem{hardy:prl13}
F. Hardy, A. E. B\"ohmer, D. Aoki, P. Burger, T. Wolf, P. Schweiss, R. Heid, P. Adelmann, Y. X. Yao, G. Kotliar, J. Schmalian, and C. Meingast
Phys. Rev. Lett. \textbf{111} 027002 (2013)

\bibitem{deMedici12}
L. de' Medici, G. Giovannetti and M. Capone, preprint at arXiv:1212.3966

\bibitem{bealmonod:prb87}
M. T. B\'eal-Monod, Phys. Rev. B \textbf{36} 8835 (1987)

\end{thebibliography}
\end{document}